\documentclass[aps,prd,preprint,tightenlines,superscriptaddress,showpacs,byrevtex]{revtex4}
\usepackage{graphicx} 
\usepackage{dcolumn}  
\usepackage{amsmath}
\usepackage{xspace}
\usepackage{SIunits}

\newcommand{\B}{\ensuremath{B}\xspace}
\newcommand{\Bz}{\ensuremath{\B^0}\xspace}
\newcommand{\BzB}{\ensuremath{\bar{\B}{}^0}\xspace}
\newcommand{\ups}{\ensuremath{\Upsilon(4S)}\xspace}
\newcommand{\jpsi}{\ensuremath{J\!/\psi}\xspace}

\newcommand{\dz}{\ensuremath{\Delta z}\xspace}
\newcommand{\dt}{\ensuremath{\Delta t}\xspace}
\newcommand{\CT}{\ensuremath{CT}\xspace}
\newcommand{\WT}{\ensuremath{WT}\xspace}

\newcommand{\taub}{\ensuremath{\tau_{\Bz}}\xspace}
\newcommand{\dm}{\ensuremath{\Delta m_d}\xspace}
\newcommand{\dg}{\ensuremath{\Delta \Gamma}\xspace}

\newcommand{\cm}{\centi\metre\xspace}
\newcommand{\micron}{\micro\metre\xspace}

\newcommand{\invfb}{\femto\reciprocal\barn\xspace}

\newcommand{\gevc}{\giga\electronvolt\!/c\xspace}
\newcommand{\gevcc}{\giga\electronvolt\!/c^2\xspace}
\newcommand{\mev}{\mega\electronvolt\xspace}

\newcommand{\mevcc}{\mega\electronvolt\!/c^2\xspace}

\graphicspath{{fig/}}

\begin{document}


\vspace*{-1.6cm}
\includegraphics[width=4cm,clip]{B-logo.epsf}

\preprint{Belle Preprint 2005-19}
\preprint{KEK Preprint 2005-17}

\title{Charge Asymmetry of Same-Sign Dileptons in \Bz-\BzB Mixing}


\affiliation{Budker Institute of Nuclear Physics, Novosibirsk}
\affiliation{Chiba University, Chiba}
\affiliation{Chonnam National University, Kwangju}
\affiliation{University of Cincinnati, Cincinnati, Ohio 45221}
\affiliation{University of Frankfurt, Frankfurt}
\affiliation{University of Hawaii, Honolulu, Hawaii 96822}
\affiliation{High Energy Accelerator Research Organization (KEK), Tsukuba}
\affiliation{Hiroshima Institute of Technology, Hiroshima}
\affiliation{Institute of High Energy Physics, Chinese Academy of Sciences, Beijing}
\affiliation{Institute of High Energy Physics, Vienna}
\affiliation{Institute for Theoretical and Experimental Physics, Moscow}
\affiliation{J. Stefan Institute, Ljubljana}
\affiliation{Kanagawa University, Yokohama}
\affiliation{Korea University, Seoul}
\affiliation{Kyungpook National University, Taegu}
\affiliation{Swiss Federal Institute of Technology of Lausanne, EPFL, Lausanne}
\affiliation{University of Ljubljana, Ljubljana}
\affiliation{University of Maribor, Maribor}
\affiliation{University of Melbourne, Victoria}
\affiliation{Nagoya University, Nagoya}
\affiliation{Nara Women's University, Nara}
\affiliation{National Central University, Chung-li}
\affiliation{National United University, Miao Li}
\affiliation{Department of Physics, National Taiwan University, Taipei}
\affiliation{H. Niewodniczanski Institute of Nuclear Physics, Krakow}
\affiliation{Nihon Dental College, Niigata}
\affiliation{Niigata University, Niigata}
\affiliation{Osaka City University, Osaka}
\affiliation{Osaka University, Osaka}
\affiliation{Panjab University, Chandigarh}
\affiliation{Peking University, Beijing}
\affiliation{Princeton University, Princeton, New Jersey 08544}
\affiliation{University of Science and Technology of China, Hefei}
\affiliation{Seoul National University, Seoul}
\affiliation{Sungkyunkwan University, Suwon}
\affiliation{University of Sydney, Sydney NSW}
\affiliation{Tata Institute of Fundamental Research, Bombay}
\affiliation{Toho University, Funabashi}
\affiliation{Tohoku Gakuin University, Tagajo}
\affiliation{Tohoku University, Sendai}
\affiliation{Department of Physics, University of Tokyo, Tokyo}
\affiliation{Tokyo Institute of Technology, Tokyo}
\affiliation{Tokyo Metropolitan University, Tokyo}
\affiliation{Tokyo University of Agriculture and Technology, Tokyo}
\affiliation{University of Tsukuba, Tsukuba}
\affiliation{Virginia Polytechnic Institute and State University, Blacksburg, Virginia 24061}
\affiliation{Yonsei University, Seoul}

   \author{E.~Nakano}\affiliation{Osaka City University, Osaka} 
   \author{K.~Abe}\affiliation{High Energy Accelerator Research Organization (KEK), Tsukuba} 
   \author{K.~Abe}\affiliation{Tohoku Gakuin University, Tagajo} 
   \author{H.~Aihara}\affiliation{Department of Physics, University of Tokyo, Tokyo} 
   \author{Y.~Asano}\affiliation{University of Tsukuba, Tsukuba} 
   \author{T.~Aushev}\affiliation{Institute for Theoretical and Experimental Physics, Moscow} 
   \author{A.~M.~Bakich}\affiliation{University of Sydney, Sydney NSW} 
   \author{Y.~Ban}\affiliation{Peking University, Beijing} 
  \author{E.~Barberio}\affiliation{University of Melbourne, Victoria} 
   \author{U.~Bitenc}\affiliation{J. Stefan Institute, Ljubljana} 
% ...
\collaboration{The Belle Collaboration}

\begin{abstract}
We report a measurement of the charge asymmetry for same-sign dileptons  
in \Bz-\BzB mixing, $A_{\rm sl}$.
The data were collected with the Belle detector at KEKB.
Using a data sample of 
78 \invfb recorded at the \ups resonance and 
9 \invfb recorded at an energy 60 \mev  below the resonance, we measure 
$A_{\rm sl} = ( -1.1 \pm 7.9(\text{stat}) \pm 7.0(\text{sys}) )\times 10^{-3}$. 
\end{abstract}

\date{\today}


\maketitle


\section{Introduction}
The Standard Model allows $CP$ violation in 
\Bz-\BzB mixing~\cite{cpvmix}.  In particular, there is a possible difference 
between the $\Bz \to \BzB$ and  $\BzB \to \Bz$ transition
rates that can manifest itself as a charge asymmetry in same-sign dilepton events in 
\ups decays when prompt leptons from
semileptonic decays of neutral \B mesons are selected. 
With the assumption of \emph{CPT} invariance, the flavor and
mass eigenstates of the neutral \B mesons are related by 
\begin{align}
|B_H\rangle &= p|\Bz\rangle + q|\BzB\rangle, \nonumber \\
|B_L\rangle &= p|\Bz\rangle - q|\BzB\rangle,
\label{eq:mixing}
\end{align}
where $|p|^2 + |q|^2 = 1$.
The time-dependent decay rate for same-sign dileptons is given by
\begin{equation}
  \Gamma_{\ups \to \ell^+ \ell^+}(\dt)
  =\frac{|A_\ell|^4}{8\taub}e^{-|\dt|/\taub}
  \left|\frac{p}{q}\right|^2
  \left[\cosh \left(\frac{\dg}{2} \dt\right) - \cos \left(\dm \dt\right)\right]
  \label{eq:rates}
\end{equation}
for the $\ell^+ \ell^+$ sample.
For the $\ell^- \ell^-$ sample, $p/q$ is replaced by $q/p$. 
Here \dm and \dg are the differences in mass and
decay width between the two mass eigenstates, \taub is
the average lifetime of the two mass eigenstates, and \dt is the
proper time difference between the two \B meson decays. In this analysis only
the absolute value of \dt is used.
It is assumed that the semileptonic decay of the neutral \B meson is
flavor specific and $CP$ conserving, so that $A_\ell = \bar{A}_\ell$,
where $A_\ell \equiv \langle X^-\ell^+\nu_\ell|B^0 \rangle$ and 
$\bar A_\ell \equiv \langle X^+\ell^-\bar\nu_\ell|{\bar B}^0 \rangle$.
If $CP$ is not conserved in mixing, the condition $|p/q| = 1$ is no
longer true and the decay rates for $\ell^+ \ell^+$ and $\ell^- \ell^-$
samples can differ. 
As can be seen in Eq.~\ref{eq:rates}, the \dt dependence is the
same for the $\ell^+ \ell^+$ and $\ell^- \ell^-$ samples, and therefore 
the $CP$ violation shows up as a \dt-independent charge asymmetry,
defined as
\begin{equation}
A_{\rm sl} \equiv 
\frac{\Gamma_{\ups \to \ell^+ \ell^+} - \Gamma_{\ups \to \ell^- \ell^-}}
     {\Gamma_{\ups \to \ell^+ \ell^+} + \Gamma_{\ups \to \ell^- \ell^-}}
 = \frac{1 - |q/p|^4}{1 + |q/p|^4}
 \simeq \frac{4 \rm{Re}(\epsilon_B)}{1 + |\epsilon_B|^2}. 
\label{eq:asl}
\end{equation}
Here $\epsilon_B$ corresponds to the $\epsilon_K$ describing
$CP$ violation in the neutral $K$ meson system. Standard Model
calculations give the size of this asymmetry to be of the order of
$10^{-3}$~\cite{cppredict,PRD}.
A significantly larger value would therefore
be an indication of new physics.

Experimentally, a measurement of same-sign dilepton events that
originate from $\Bz\Bz$ and $\BzB\BzB$ initial states
requires careful charge-dependent corrections, which are done in
several steps. First, the contribution from continuum $e^+ e^- \to q\bar q$
(where $q=u,d,s$ or $c$) to same-sign dilepton events is
subtracted using off-resonance data. Second, all detected lepton 
tracks are corrected for charge asymmetries in
the efficiencies for track finding and 
lepton identification, and for 
the probabilities of misidentifying hadrons as leptons.
After these corrections, 
the remaining same-sign dilepton events still 
contain backgrounds from $\Bz\BzB$ and $\B^+ \B^-$ events.
The last step of this analysis is to separate the
signal events from these background events using their different behavior
in the \dt distributions. 

\section{Belle Detector}
This analysis is based on a data sample of 78 \invfb at the \ups resonance
(``on-resonance'') and 9 \invfb at 60 MeV below
the \ups resonance (``off-resonance'') collected with 
the Belle detector~\cite{belle} at the KEKB
asymmetric $e^+e^-$ collider~\cite{kekb}.
The Belle detector is a large-solid-angle magnetic
spectrometer that 
consists of a three-layer silicon vertex detector (SVD),
a 50-layer central drift chamber (CDC) for tracking, a mosaic of
aerogel threshold Cherenkov counters, time-of-flight
scintillation counters (TOF), and an array of CsI(Tl) crystals
for electromagnetic calorimetry (ECL)  located inside of
a superconducting solenoid coil that provides a 1.5~T
magnetic field.  An iron flux-return located outside of
the coil is instrumented to detect $K_L^0$ mesons and to identify
muons (KLM).  

\subsection{Track Finding Efficiency}
\begin{figure}[thb]
\includegraphics[width=17cm,clip]{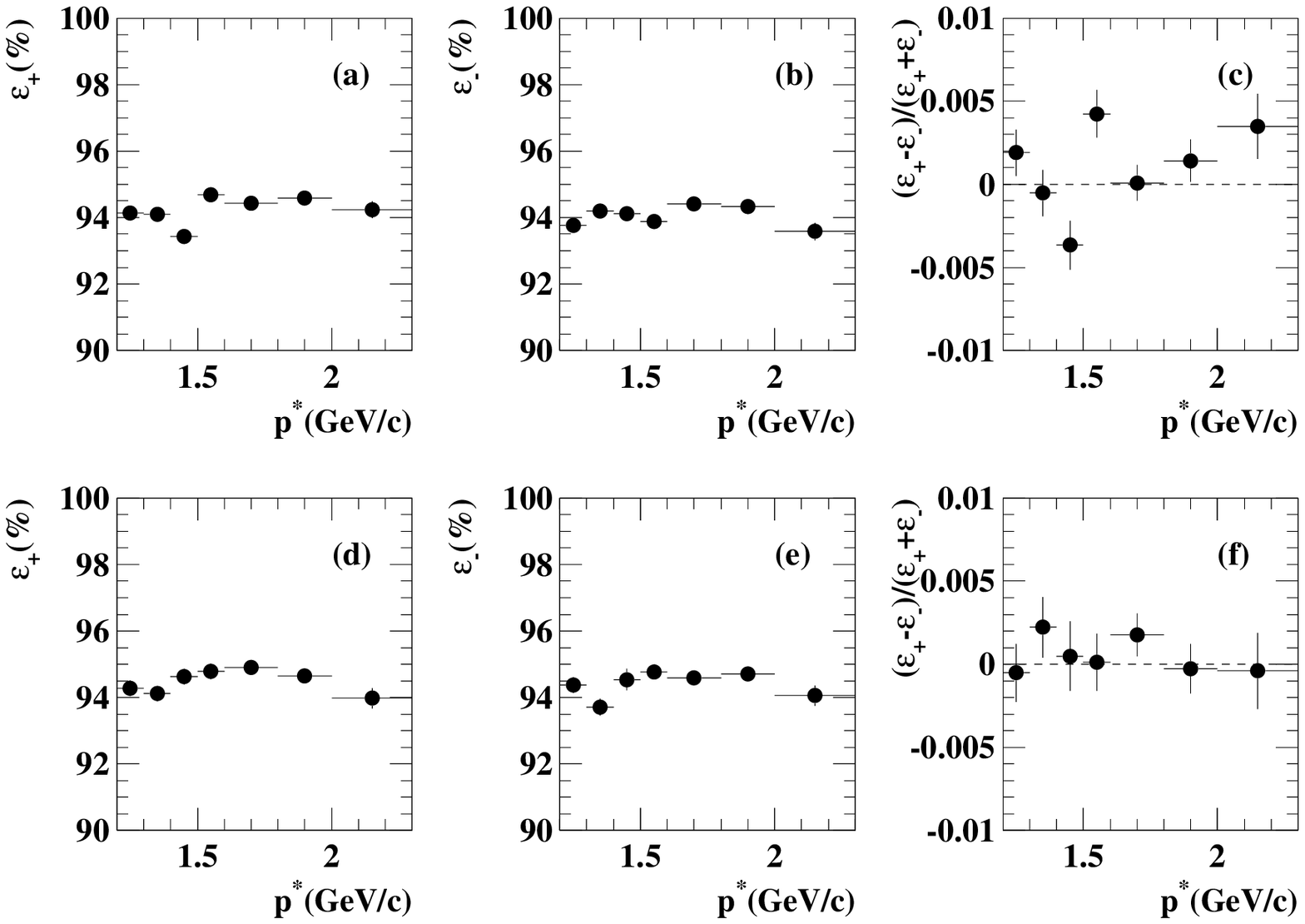}
\caption{$\theta_\text{lab}$-averated track finding efficiencies as a function of c.m. momentum
 for positron  tracks $\varepsilon_+$ (a), electron tracks
 $\varepsilon_-$ (b), and charge dependence defined as 
 $(\varepsilon_+ - \varepsilon_-)/(\varepsilon_+ + \varepsilon_-)$ (c).
 Corresponding plots for muon tracks are shown in (d), (e), and (f)}.
\label{seleffe}
\end{figure}

The track finding efficiency is determined by analyzing a sample
where simulated single electron or muon tracks are overlaid on
hadronic events taken from experimental data. Lepton tracks are
generated to cover the region of $1.2~\gevc <p^*<2.3~\gevc$ and
$30^{\circ}<\theta_\text{lab}< 135^{\circ}$, 
where $p^*$ is the lepton momentum in the $e^+ e^-$ center-of-mass (c.m.)
frame and $\theta_\text{lab}$ is the
angle of the lepton track with respect to the $z$-axis in the laboratory
frame. The $z$-axis passes through the nominal interaction point, and
is anti-parallel to the positron beam direction.
The track finding efficiencies are obtained for each bin of
$p^*$
(1.2--1.3, 1.3--1.4, 1.4--1.5,
1.5--1.6, 1.6--1.8, 1.8--2.0
and 2.0--2.3~$\gevc$),
and 
$\theta_{\text{lab}}$
(30$^\circ$--37$^\circ$, 37$^\circ$--50$^\circ$,
50$^\circ$--77$^\circ$, 77$^\circ$--82$^\circ$,
82$^\circ$--111$^\circ$, 111$^\circ$--119$^\circ$,
119$^\circ$--128$^\circ$ and 128$^\circ$--135$^\circ$).
Figure~\ref{seleffe}
shows $\theta_\text{lab}$-averaged track finding efficiencies for
positive and negative tracks
separately and their fractional differences as  functions of $p^*$ for
electron and muon tracks. Events in all $\theta_\text{lab}$ regions
are combined in these plots. The charge dependence of the track finding
efficiency for both electrons and muons is less than 1.0\%.

\subsection{Lepton Identification}
The most important contribution to the electron identification comes
from examination of the ratio of the ECL cluster energy to the track
momentum measured in the CDC. This information is combined with the
shower measurement in the ECL, the specific ionization measurements
($dE/dx$) in the CDC,
matching information between the ECL cluster position and extrapolated position
of the CDC track, and the ACC light yield to form an electron likelihood
${\cal L}_e$~\cite{eid}.

The two-photon process $e^+e^- \to (e^+e^-)e^+e^-$ is used to 
estimate the electron identification efficiency.
For this data sample, events are required to have: i) 
two tracks with particle identification information inconsistent
with a muon hypothesis,
laboratory momenta greater than $0.5~\gevc$ and
transverse momenta greater than $0.25~\gevc$;
ii) at least one ECL cluster with energy greater than $20~\mathrm{MeV}$. 
The two tracks are required to have:
i) an acolinearity angle whose cosine is greater than $-0.997$;
ii) a transverse-momentum sum less than $0.2~\gevc$;
iii) a longitudinal momentum sum of less than $2.5~\gevc$ in the c.m. frame;
iv) an invariant mass less than $5~\gevcc$.  In addition, the sum of the 
ECL cluster energies has to be between 0.6 GeV and 6.0 GeV.
The electron identification efficiency is obtained by taking the ratio of 
the number of tracks selected with the above requirements with and without 
additional electron identification requirements.

For muon identification, CDC tracks are extrapolated to the KLM and 
the measured range and transverse deviation in the KLM is compared with 
the expected values to form a muon likelihood $\mathcal{L}_\mu$~\cite{muonid}.
The muon identification efficiency is determined by analyzing a data sample 
where simulated single-muon tracks are overlaid on the hadronic events 
taken from experimental data. 

\begin{figure}[!thb]
\includegraphics[width=17cm,clip]{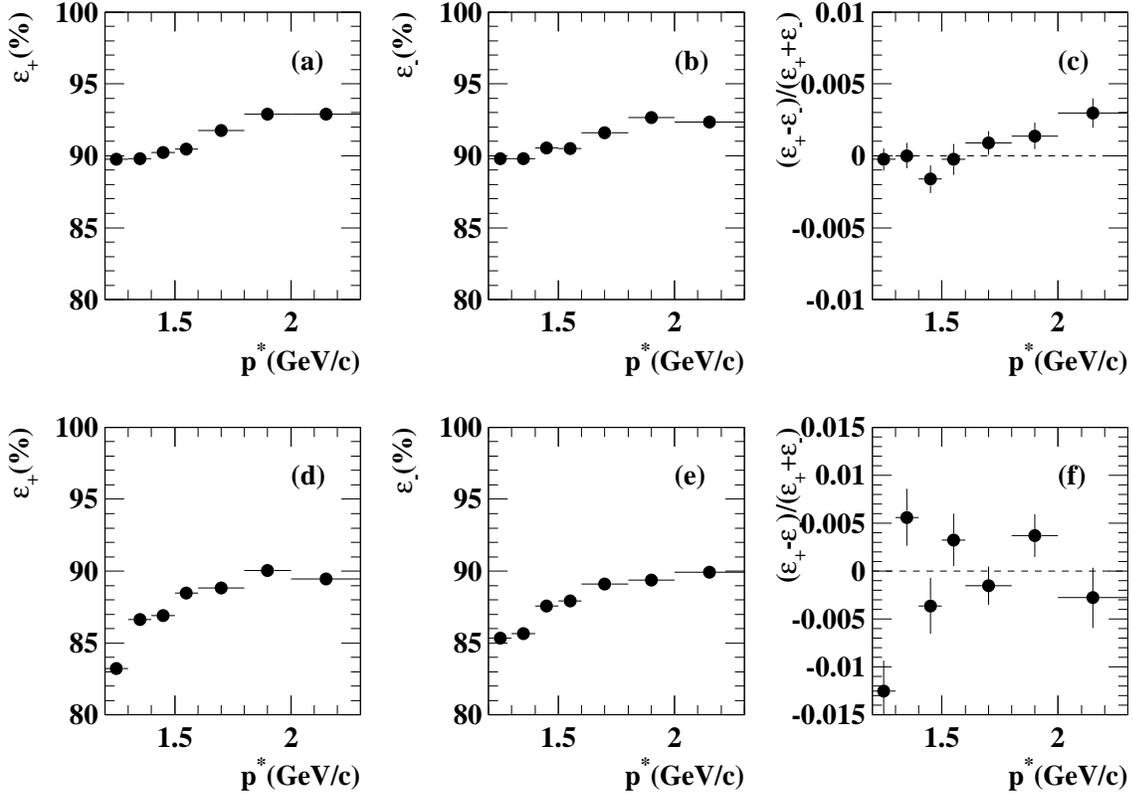}
\caption{The $\theta_{\text{lab}}$-averaged identification efficiencies
as a function of c.m. momentum 
for positrons $\varepsilon_+$ (a), electrons $\varepsilon_-$ (b), 
  and charge dependence defined as
 $(\varepsilon_+-\varepsilon_-)/(\varepsilon_+ + \varepsilon_-)$ (c).
Corresponding plots 
 for muons are shown in (d), (e), and (f). 
}
\label{leptonid}
\end{figure}

The lepton identification efficiencies are obtained in the same 
$p^*$ and $\theta_{\text{lab}}$ bins used for the track finding efficiency
study.
Figure~\ref{leptonid} shows the $\theta_{\text{lab}}$-averaged
charge-dependent lepton identification
efficiencies, where electron tracks are required to satisfy
$\mathcal{L}_e > 0.8$, and the muon tracks are required to satisfy
$\mathcal{L}_\mu > 0.9$ and to have a reduced $\chi^2$
value for their transverse
deviations in the KLM of less than 3.5.
The charge
dependence of the identification efficiency is less than
1\% for both electrons and muons.
Exactly the same criteria are imposed to select signal leptons.

\subsection{Hadron Misidentification}
The information on the ACC light yield,
the time-of-flight measurement, and $dE/dx$ 
measurements in the CDC 
is combined to provide hadron likelihoods: $\mathcal{L}_\pi$ for pions, 
$\mathcal{L}_K$ for kaons, and $\mathcal{L}_p$ for protons. 

The hadron fake rate, which is defined as the probability that a hadron track 
is mis-identified as a lepton, is determined from a sample of 
$K^0_S \to \pi^+\pi^-$ decays for pions, 
$\phi \to K^+K^-$ decays for kaons, and $\Lambda \to p\pi^-$ 
($\bar{\Lambda} \to \bar{p} \pi^+$) decays for protons.
These decays are selected from a hadronic event sample described below.
To select these track-pair combinations, the distance of closest approach with
respect to the run-dependent interaction point and the position of the decay
vertex are used. The difference in $z$ position of the two tracks and the
angle between position vector of the decay vertex and the momentum vector
of $K^0_S$ or $\Lambda$($\bar\Lambda$) in $r$-$\phi$ plane
at the decay vertex are also used.
When evaluating the fake rate for positively charged tracks,
invariant masses are plotted for a sample of pairs where both 
tracks satisfy the above criteria and the negative track satisfies 
an additional hadron identification requirement.
The signal yields are obtained by fitting the resulting
mass distribution to a sum of a 
double Gaussian signal and a smooth background function in two ways:
once without imposing any lepton identification requirement and
once after imposing the lepton identification requirement on the positive
track.  The ratio of the two signal yields gives
the fake rate for the positive charged tracks.
The following requirements are imposed on the likelihood ratios:
$\mathcal{L}_\pi/(\mathcal{L}_\pi + \mathcal{L}_K)> 0.8$ for pions in 
$K^0_S \to \pi^+ \pi^-$, 
$\mathcal{L}_K/(\mathcal{L}_K+ \mathcal{L}_\pi)> 0.8$ for kaons in 
$\phi \to K^+ K^-$, and 
$\mathcal{L}_\pi/(\mathcal{L}_\pi + \mathcal{L}_p)> 0.8$ for pions in 
$\Lambda \to p \pi^-$.
The fake rate for negatively charged tracks is obtained by repeating the 
procedure above with the roles of negative and positive tracks reversed.  

In this study, because of low statistics, $\theta_{\text{lab}}$-averaged
values are used, and some $p^*$ bins are combined.

\begin{figure}[!thb]
\includegraphics[width=17cm,clip]{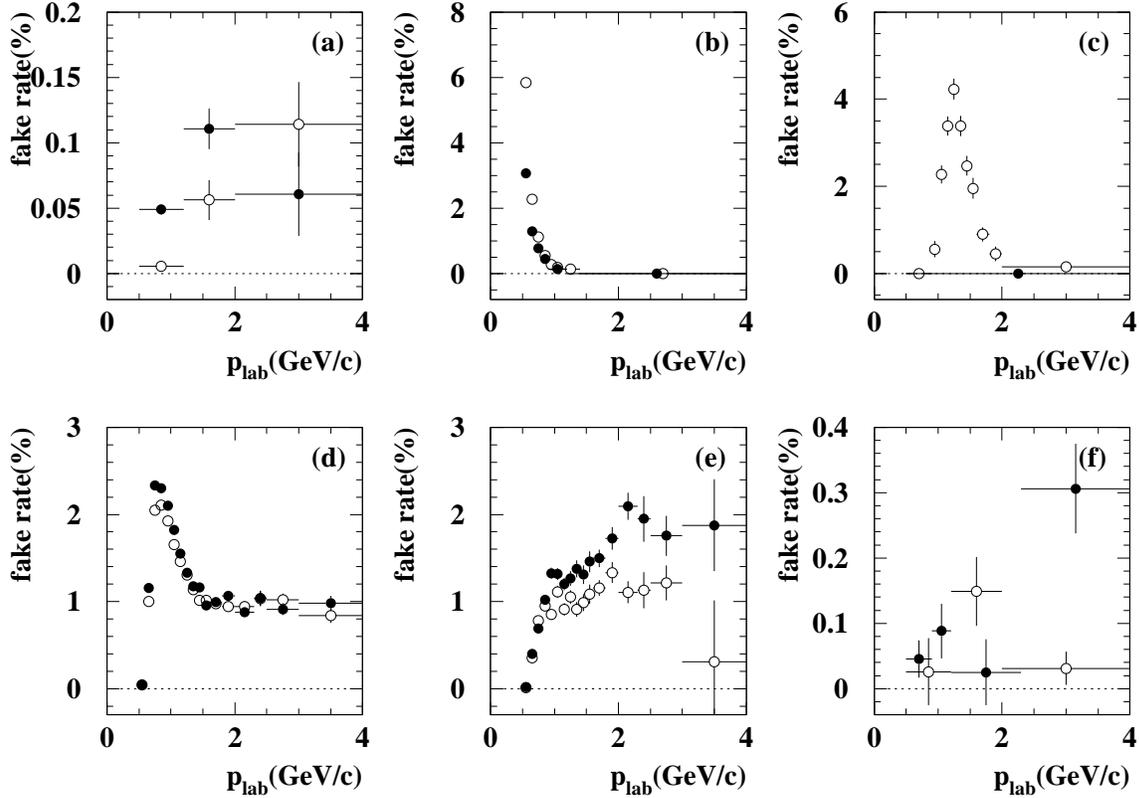}
\caption{
  Rates of pions ((a) and (d)), kaons ((b) and (e)), and protons ((c)
  and (f)) faking electrons and muons vs. laboratory momentum.  Filled
  circles are for positive tracks and open circles are for negative tracks.
  The increase in the rate of kaons faking electrons at low momentum
  clearly visible in (b) is due to the overlap of the electron and
  kaon energy-loss bands; the other distributions are discussed in the text.
  Note that the minimum values of vartical axes are set to negative values.
  The dotted lines shows fake rate equal to zero.}
\label{fake}
\end{figure}

Figure~\ref{fake} shows the hadron fake rates
as a function of momentum in the laboratory frame.
The rate of pions faking electrons is at most 0.1\% for both charges.
The rate of kaons faking
electrons decreases rapidly as $p_\text{lab}$ becomes larger and is
less than 0.2\% for $p_\text{lab} > 1.4~\gevc$, with no significant
charge dependence. While the rate of protons faking positrons is
nearly zero, the rate for anti-protons faking electrons is
as large as 4\% due to the large
anti-proton annihilation cross section in the ECL.
Because of low statistics, the rate of protons faking positrons is 
obtained from the $\theta_{\text{lab}}$-averaged value over all momenta.
The rate of pions
faking muons is about 1\% for $p_\text{lab} > 1.5~\gevc$ and shows
no significant charge dependence. The rate of kaons faking muons is 1\%
to 2\%, with that for $K^+$ being about 50\% larger than that for $K^-$
due to the
larger kaon-nucleon cross section for the $K^-$. The rate of protons faking
muons is less than 0.4\% and shows no clear charge dependence.

\section{Event Selection}

\subsection{Hadronic Event Selection}
Hadronic events are required to have at least five tracks, an event
vertex with radial and $z$ coordinates within $1.5~\cm$ and $3.5~\cm$,
respectively, of the nominal beam interaction point, a total
reconstructed c.m. energy greater than $0.5~W$ ($W$ is the c.m.
energy), a net reconstructed c.m. momentum with a $z$ component less
than 0.3~$W$$/c$, a total energy deposited in the ECL between $0.025$ and
$0.9~W$, and a ratio $R_2$ of the second and zeroth Fox-Wolfram
moments~\cite{Fox-Wolfram} of less than 0.7.

\subsection{Dilepton Event Selection}
Lepton candidates are selected from among the charged tracks
by requiring the criteria previously described.
In both electron and muon cases, a distance of closest approach to
the run-dependent interaction
point less than $0.05~\cm$ radially and $2.0~\cm$ in $z$ is required.  At least
one SVD hit per track in the $r$-$\phi$ view and two SVD hits in the $r$-$z$ view are required.
To eliminate electrons from $\gamma \to e^+ e^-$
conversions, electron candidates are paired with all other oppositely
charged tracks and the invariant mass (assuming the electron mass)
$M_{e^+e^-}$ is calculated. If $M_{e^+e^-} < 100~\mevcc$,
the electron candidate is rejected. If a hadronic event contains more
than two lepton candidates, the two with the highest c.m. momenta are
used.

The two lepton candidates must satisfy additional criteria.  
The c.m. momentum of each lepton is required to be in the range 
$1.2~\gevc < p^* < 2.3~\gevc$. The lower threshold reduces contributions from
secondary charm decay; the upper threshold reduces continuum
contributions.
Each lepton track must be in the range
$30^\circ <\theta_\text{lab}<135^\circ$, where it has
better $z$ vertex resolution and lepton identification.
Events that contain one or more \jpsi candidates are rejected.
The invariant mass of each candidate lepton paired with each
oppositely charged track (assuming the corresponding lepton mass)
is calculated. If the invariant mass falls within the \jpsi region,
defined as $-0.15~\gevcc < ( M_{e^+ e^-} - M_{\jpsi})< 0.05~\gevcc$ or 
$-0.05~\gevcc < (M_{\mu^+\mu^-} - M_{\jpsi})< 0.05~\gevcc$,
the candidate event is rejected. The looser lower mass window for
the electron pair invariant mass accomodates bremsstrahlung
of the daughter electron(s).

\begin{figure}[!thb]
\includegraphics[width=17cm,clip]{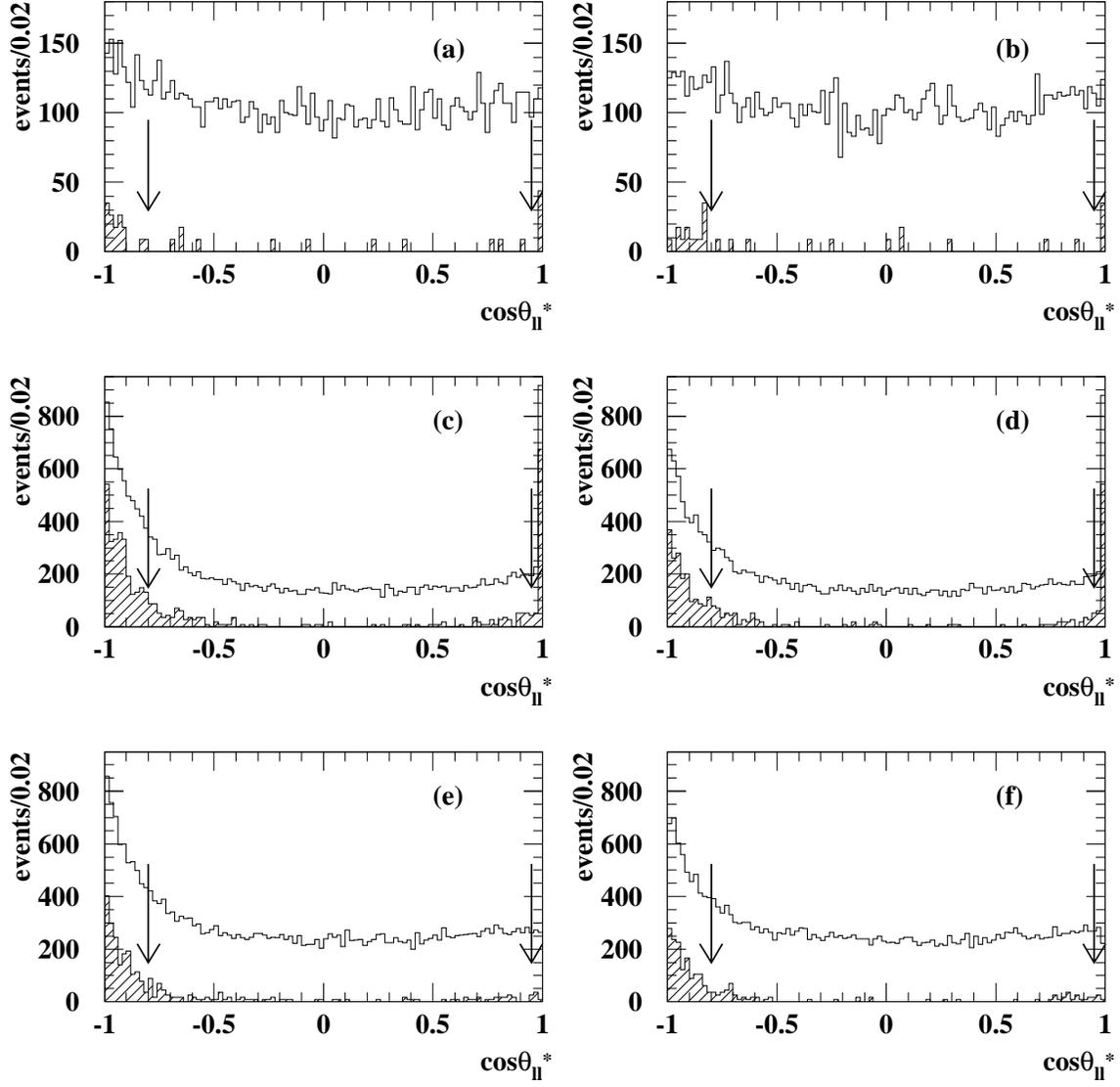}
\caption{
$\cos \theta^*_{\ell\ell}$ distributions for dilepton
samples in the on-resonance (open histogram) and scaled
off-resonance (filled histogram) data.
(a),(b) show $ee$ events, (c),(d) show $\mu\mu$ events and (e),(f) are
from $e\mu$ combinations. (a),(c) and (e) are the $++$ charge case and
(b), (d) and (f) are the $--$ charge case.
The arrows indicate the selection requirements.
}
\label{openang}
\end{figure}

As can be seen in Figure~\ref{openang}, distributions of the opening
angle of the two tracks in the c.m. frame, $\cos \theta^*_{\ell\ell}$,
for the $\mu \mu$ and $e\mu$ pairs show distinct peaks in the 
back-to-back direction ($\cos\theta^*_{\ell\ell}\simeq -1$).
This background is caused by hadron tracks misidentified as muons
among jet-like continuum events.
Also, spikes can be seen at $\cos\theta^*_{\ell\ell} = 1$ for the $\mu\mu$
pair events.
This structure is caused by jet-like continuum events where a non-muon
track is identified as a muon because hits in the KLM from a nearby
true muon are assigned to it.
The dilepton opening angle in the
c.m. frame $\theta^*_{\ell\ell}$ is required to satisfy 
$-0.80 < \cos\theta^*_{\ell\ell} < 0.95$ in order to reduce this background.

With these selection criteria there are 46533 positive and 45477
negative same sign dilepton events found in the on-resonance data.
Continuum contributions are estimated to be 2230 for positive and
1574 for negative same sign events, based on the yield from
off-resonance data. To estimate the continuum contribution from off-resonance
data, off-resonance yields were scaled by the integrated luminosities and
cross-section ratio. The scaling factor is defined by
\begin{equation}
f = \frac{\int \mathcal{L}_{\text{on}} dt}
     {\int \mathcal{L}_{\text{off}} dt}
\frac{s_{\text{off}}}
     {s_{\text{on}}}. 
\label{eq:sf}
\end{equation}
where $\int \mathcal{L}_{\rm{on}(\rm{off})} dt$ and
$s_{\rm{on}(\rm{off})}$ are 
the integrated luminosities and the square of c.m. energies for
on(off)-resonance, respectively.

These dilepton yields, decomposed into three lepton
categories, are given in Table~\ref{num}.

\begin{table}[!htb]
  \begin{center}
    \caption{Numbers of dilepton events.}
    \begin{tabular}{c|r|r||r|r||r|r}
      \hline
      \hline
      & \multicolumn{2}{c||}{On-resonance}&\multicolumn{2}{c||}{Off-resonance}&\multicolumn{2}{c}{Continuum}\\
      Combination & \multicolumn{1}{c|}{positive} & \multicolumn{1}{c||}{negative} & \multicolumn{1}{c|}{positive} & \multicolumn{1}{c||}{negative} & \multicolumn{1}{c|}{positive} & \multicolumn{1}{c}{negative} \\
      \hline
      $ee$        &  9059 &  9028 &  11 &  11 &   96.2$\pm$ 28.9 &  96.2$\pm$ 28.9 \\
      $\mu\mu$    & 14672 & 14014 & 144 & 100 & 1259.2$\pm$104.9 & 874.4$\pm$ 87.4 \\
      $e\mu$      & 22802 & 22435 & 100 &  69 &  874.4$\pm$ 87.4 & 603.4$\pm$ 72.6 \\
      \hline
      total       & 46533 & 45477 & 255 & 180 & 2229.8$\pm$139.6 &1574.0$\pm$117.3 \\
      \hline
      \hline
    \end{tabular}
    \label{num}
  \end{center}
\end{table}

\subsection{{\boldmath \dz} Determination}
The $z$-coordinate of each \B meson decay vertex is the production 
point of the daughter lepton, which is determined from 
the intersection of the lepton track with the run-dependent profile of 
the interaction point.
The distance between 
the $z$-coordinates of the two leptons,
$ |\Delta z| $, is defined as $ |\Delta z| =  |z(\ell_1) - z(\ell_2)|$.

In order to estimate the effect of detector resolution in the \dz distribution,
\jpsi decays to $e^+e^-$ and $\mu^+\mu^-$ are used.
In these events,
the two tracks originate from the same point, so the measured \dz,
after the background contribution is subtracted, yields the detector
resolution. Candidate \jpsi mesons are selected using the same requirements
used for dilepton events, except for 
the \jpsi veto.  The \jpsi signal regions
are defined as $3.00~\gevcc < M(e^+e^-) < 3.14~\gevcc$ and $3.05~\gevcc
< M(\mu^+\mu^-) < 3.14~\gevcc$ and the sideband region as $3.18~\gevcc
< M(\ell^+\ell^-) < 3.50~\gevcc$ for both electrons and muons.

The invariant mass distributions of $J/\psi$ candidates are fitted to a
function given by
\begin{equation}
N(M) = h_0 e^{ -\frac{(M - M_0)^2}{2S^2 } }
+ h_1 e^{ -\frac{(M - M_0)^2}{2 {\sigma_1}^2} }
+ A(M-B)^2 + C.
\end{equation}
Here, $h_0$ and $h_1$ are the normalizations of the two Gaussians used to
describe the signal, $M_0$ is
the Gaussian mean which is common to both Gaussians, and
$\sigma_0$ and $\sigma_1$ are the standard deviations of the Gaussians.
A parameter $S$, defined as $S = \sigma_0$ for $M \ge M_0$ and
$S = \sigma_0 + \alpha(M - M_0)$ for $M < M_0$,
is introduced to modify the lower mass tail of one of the Gaussians
to take the effect of bremsstrahlung into account.
$A$, $B$ and $C$ are the parameters of the background function.
The
\dz distribution of the sideband region is scaled to the background
yield in the signal region and subtracted from the signal region \dz
distribution.

\begin{figure}[!thb]
\includegraphics[width=17cm,clip]{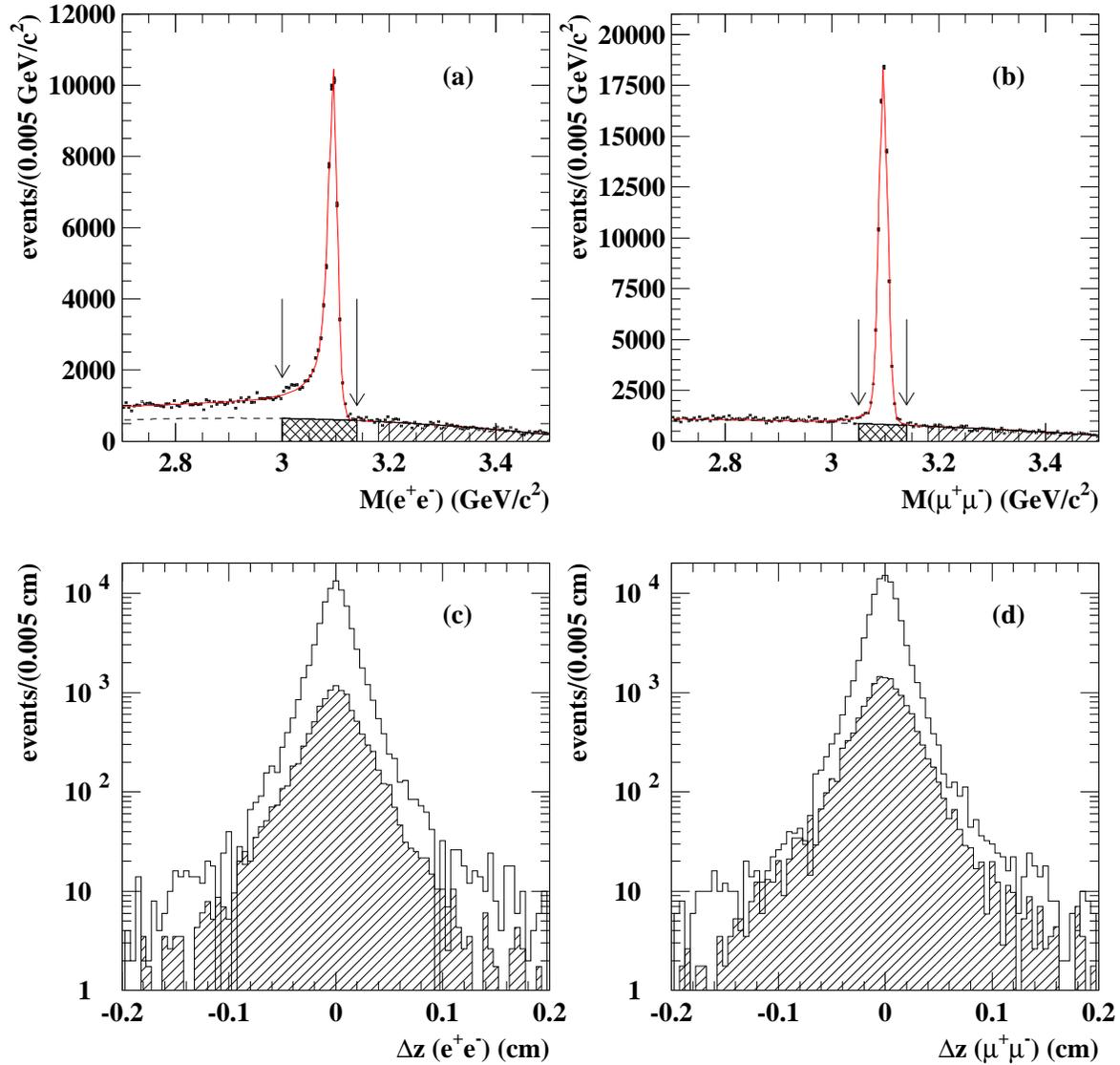}
\caption{Mass distributions for $\jpsi \to e^+e^-$ (a) and $\jpsi \to 
  \mu^+\mu^-$ (b). The arrows indicate the selection criteria for each
  decay mode.  The dashed lines indicate the fitted background
  component and the solid lines show the total fit results.  The
  cross hatched area shows the estimated combinatorial background in
  the signal region and the single hatched area is the sideband region
  used to estimate the background \dz distribution. The \dz distributions
  for $\jpsi \to e^+e^-$ (c) and $\jpsi \to \mu^+\mu^-$ (d). Open
  histograms are for all \jpsi candidates in the signal region and
  hatched histograms are for the background.}
\label{mrjpsi}
\end{figure}

The \jpsi mass distributions and the \dz distributions are 
shown in Figure~\ref{mrjpsi}.
The RMS values of the \dz distributions are 193 $\micron$ for
$\jpsi \to e^+ e^-$, $177~\micron$
for $\jpsi \to \mu^+ \mu^-$, and $185~\micron$ for the combined
$\jpsi \to \ell^+\ell^-$ sample.

\subsection{Subtraction of Continuum Events}
A sample of dilepton events originating from $B\bar{B}$ events is obtained by
subtracting the luminosity and cross-section scaled off-resonance 
data from the on-resonance data. 
Since the kinematics of dilepton candidates in $B\bar{B}$ decays is different
from those in continuum events in each of the variables
($p^*_1$, $p^*_2$, $\theta^*_1$, $\theta^*_2$, $\theta^*_{\ell\ell}$, \dz),
where $\theta^*_{1(2)}$ is the polar angle
in the c.m. frame with respect to the beam axis of the more
(less) energetic lepton, the subtraction should, in principle, be
performed in this six-dimensional space, separately for each lepton flavor
and charge combination.

\begin{figure}[!thb]
\includegraphics[width=17cm,clip]{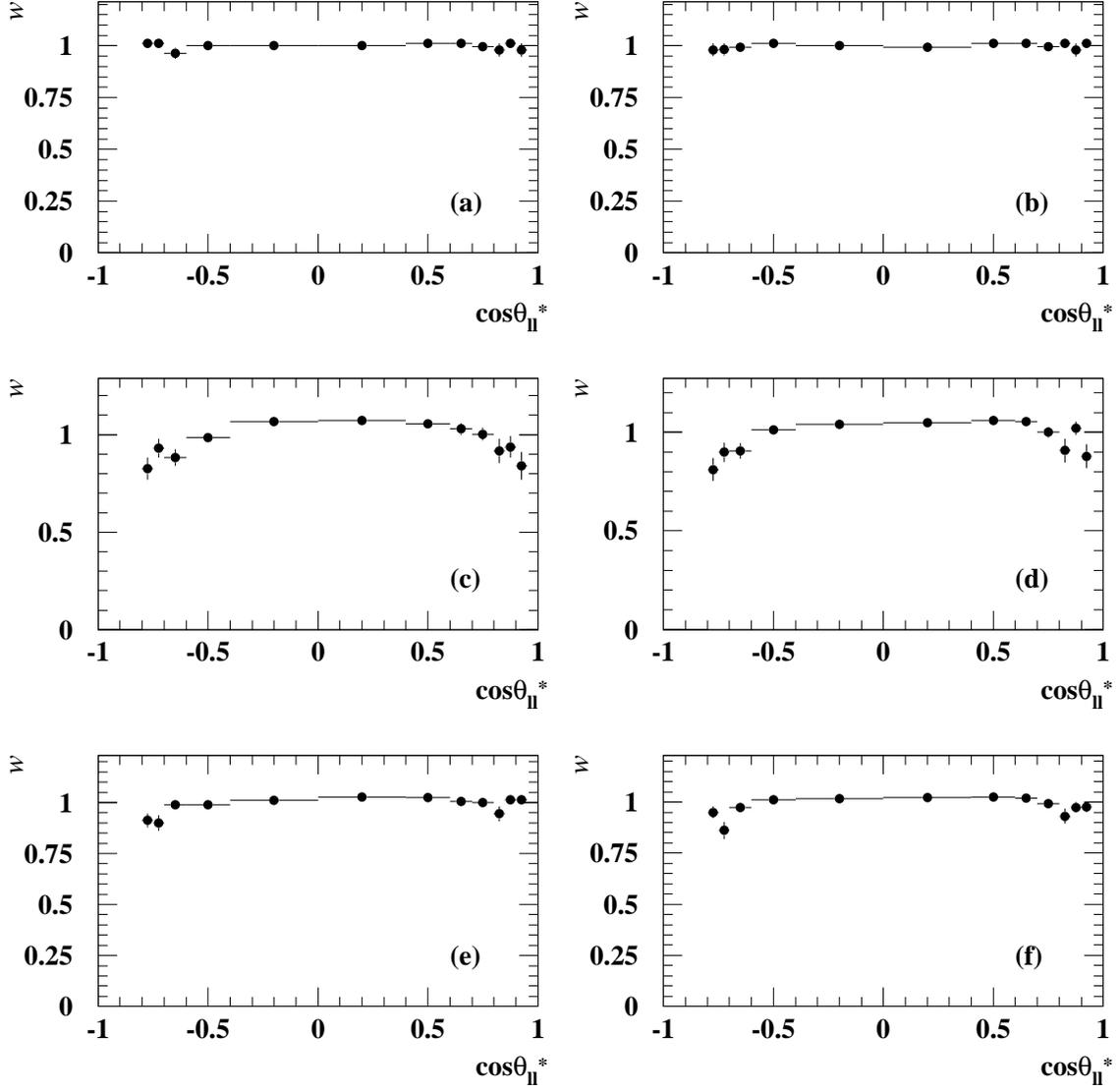}
 \caption{$\cos \theta^*_{\ell\ell}$ dependence of the weighting factor for the 
fraction of the dilepton yield originating from $B \bar B$ decays in the on-resonance 
 data for $e^+e^+$ (a) and $e^-e^-$ (b) and
corresponding quantities for $\mu\mu$ (c) and (d) and $e\mu$ (e) and (f).}
\label{wopang}
\end{figure}

Given the available statistics, this approach is not possible. Instead, we perform
the subtraction by weighting the on-resonance and off-resonance yields
for one of the six kinematical variables, while integrating over the five other variables.
We obtain weighting factors for the six variables by
repeating this procedure. Since, to a first approximation, the six
variables are not correlated with each other, this approach provides
the $B \bar B$ yield in the six variable space.
The weighting factors  are given by
$w(k) =(1/r_{BB})
  (N_{\mathrm{on}}(k) -f N_{\mathrm{off}}(k))/N_{\mathrm{on}}(k)$
where $k$ denotes each of six variables,
$f$ is the scaling factor for the luminosity and c.m. energy introduced
in Eq.~\ref{eq:sf},
and $r_{BB} \equiv N_{B \bar B}^\mathrm{total}/ N_\mathrm{on}^\mathrm{total}$
is the fraction of total $B \bar B$ events in the on-resonance yield
after integrating over
all six variables; it is used for normalization.
While the weighting
factors show very little dependence on $p^*_1$, $p^*_2$, $\theta^*_1$,
and $\theta^*_2$ for all combinations of lepton flavors and charges,
a clear dependence is observed for $\theta^*_{\ell\ell}$ in the case
of the $\mu \mu$ and $e\mu$ data samples as shown in Figure~\ref{wopang}.
A clear dependence on $\Delta z$ is also seen for all lepton pair
combinations.

Using this method, the dilepton candidate yield for each 
lepton flavor and charge combination is then given 
in terms of the on-resonance yield and the weighting factors
by 

\begin{equation}
N_{B\bar{B}}(p^*_1, p^*_2, \theta^*_1, \theta^*_2, \theta^*_{\ell\ell}, \dz ) 
= r_{BB}  \prod_{k} w(k) 
N_{\rm{on}}( p^*_1, p^*_2, \theta^*_1, \theta^*_2, \theta^*_{\ell\ell}, \dz ).
\label{eq:nbbbar}
\end{equation}
The \dz distributions of the dilepton yields
are obtained by projecting
$N_{B\bar{B}}(p^*_1, p^*_2, \theta^*_1, \theta^*_2, \theta^*_{\ell\ell}, \dz)$
onto the $|\dz|$ axis.

\section{Result}

\subsection{Corrections to Lepton Candidates}
\begin{figure}[!thb]
\includegraphics[width=17cm,clip]{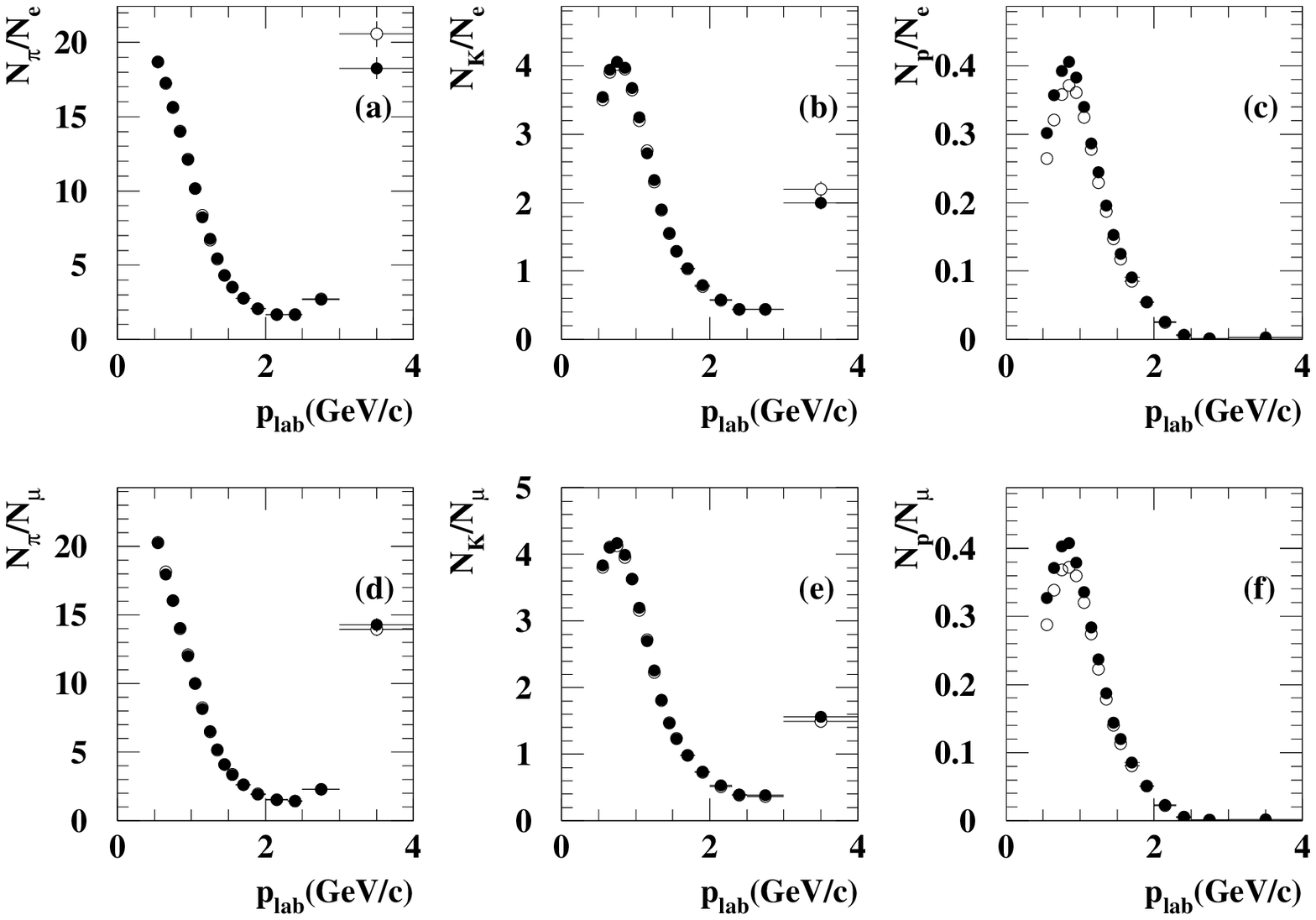}
\caption{Relative multiplicities of hadrons as a function of $p_\text{lab}$
  with respect to leptons. (a), (b) and (c) are for electrons and (d), (e) and
  (f) are for muons. (a) and (d) are pion to lepton, (b) and (e) are kaon to
  lepton, (c) and (f) are proton to lepton ratios.
  Filled circles are for positive
  tracks and open circles are for negative tracks. The effects of
  decay-in-flight and interaction with detector are included.
  Though in (a),
  (b), (d) and (e), the difference between positive and negative cases is
  less than 1\%, in both (c) and (f), the proton rate is larger than
  the anti-proton rate by about 8\%.
  The increases in multiplicities for pions and kaons around 3.5 \gevc are
  due to two-body hadronic $B$ decays.}
\label{hlratio}
\end{figure}

The number of detected leptons for each lepton flavor and charge
$N^\pm_{\text{det}}$ is related to the number of true leptons 
$N^\pm_\ell$ by
\begin{eqnarray}
  N^\pm_{\text{det}}(p^*,\theta_{\text{lab}}) =
  N^\pm_\ell(p^*,\theta_{\text{lab}})
  \varepsilon^\pm_{\text{trk}}(p^*,\theta_{\text{lab}})
  \left\{\varepsilon^\pm_{\text{pid}}(p^*,\theta_{\text{lab}})
  + \sum_{h=\pi,K,p} 
r^\pm_{h\ell}(p^*, \theta_{\text{lab}})
\eta^\pm_{h\ell}(p^*, \theta_{\text{lab}}
 )\right\},
\label{ndet}
\end{eqnarray}
where $\varepsilon^\pm_{\text{trk}}$ and $\varepsilon^\pm_{\text{pid}}$ are
the efficiencies for track finding and lepton identification, 
$r^\pm_{h \ell}$ is the relative multiplicity of hadron of type   
$h$ with respect to leptons of type  $\ell$ in $B \bar B$ event, 
and $\eta^\pm_{h \ell}$ is the rate of hadrons $h$ faking  leptons $\ell$. 
The relative multiplicities are determined from $B \bar B$ 
Monte Carlo (MC) events, and are shown in Figure~\ref{hlratio}.
The relative multiplicities include the effects of
decay-in-flight and interaction
with the detector.

Using the measured efficiencies, fake rates and the MC-determined relative
multiplicities, the correction factors
$N_\ell/N_{\text{det}}$ are determined in 7 bins of $p^*$   
(1.2--1.3, 1.3--1.4, 1.4--1.5,
1.5--1.6, 1.6--1.8, 1.8--2.0
and 2.0--2.3~$\gevc$),
and 
8 bins of $\theta_{\text{lab}}$ 
(30$^\circ$--37$^\circ$, 37$^\circ$--50$^\circ$,
50$^\circ$--77$^\circ$, 77$^\circ$--82$^\circ$,
82$^\circ$--111$^\circ$, 111$^\circ$--119$^\circ$,
119$^\circ$--128$^\circ$ and 128$^\circ$--135$^\circ$). 
(The fake rates are measured in the laboratory frame, but are converted
into $p^*$- and $\theta_{\text{lab}}$-dependent measurements for this
correction.)

After the correction, the dilepton sample contains true leptons that come 
either from prompt neutral \B meson decay (signal) or from background
processes such as charged \B meson decay, secondary charm decay, or
other leptonic \B meson processes.

\subsection{Fit to {\boldmath \dz} Distribution}
A binned maximum likelihood fit with signal and background
contributions is used to extract $A_{\rm sl}(|\dz|)$ from the \dz
distribution. 
The overall background level is obtained from a fit with positive
($++$) and negative ($--$) samples combined. In this fit the \dz distribution
for signal events is given by Eq.~\ref{eq:rates},
assuming $\dg$ is small~\cite{PDG},
\begin{equation}
  P^{\text{SS}} \propto e^{-|\dt|/\taub}(1-\cos(\dm \dt)),
  \label{pmix}
\end{equation}
convolved with the detector response function described earlier. 
Here, \taub  and \dm are fixed to their world average values~\cite{PDG}.

\begin{figure}[!thb]
\includegraphics[width=17cm,clip]{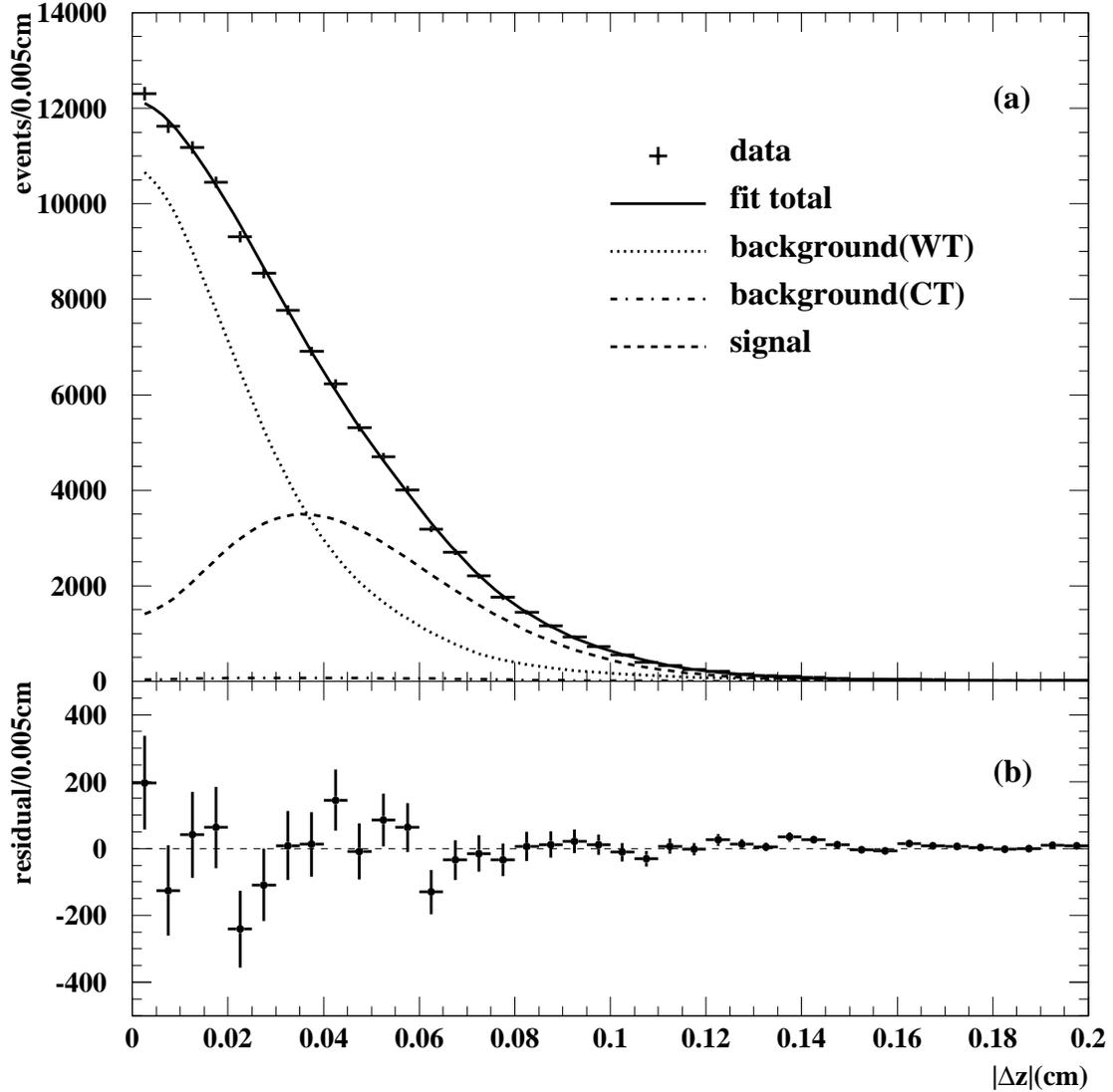}
 \caption{(a) \dz distribution for the true dilepton
   events ($++$ and $--$ are combined). 
   Points with error bars are data. The dot-dashed line shows the
   contribution from \CT backgrounds, the dotted line shows the \WT
   background contributions, the dashed line indicates the signal
   component and the solid line indicates the fitted total.
   (b) Difference between data and fit result as a function
   of $|\Delta z|$.} 
\label{dz}
\end{figure}

The backgrounds are divided into two categories: correctly tagged
(\CT), and wrongly tagged (\WT).
The \CT category mainly contains events in which both leptons
come from secondary charm decay in
$\Bz\BzB \to \Bz\Bz
(\BzB\BzB)$ (mixed) processes. The \WT category contains events in
which one lepton is from secondary charm decay of unmixed $\Bz\BzB$ or
$B^+ B^-$
and the other is from a semileptonic $B$ decay.
Though background \dz distributions are estimated using MC
simulations, the MC underestimates the width of the \dz distribution.
To correct for this, the MC \dz
distribution is convolved with a Gaussian of standard deviation
$\sigma = 69~\micron$
~\cite{hastings}.
The \dz distribution for the true same-sign dilepton events,
where positive ($++$) and negative ($--$) samples are combined,
is shown 
in Figure~\ref{dz}-(a) together with the fit results. 
The $\chi^2/n.d.f.$ of the fit is 48.75/38
(note that only statistical errors are included).
In the fit, the ratio of \CT to \WT is fixed at the MC value, and
only the ratio of signal and background is allowed to float.
The MC-estimated \CT and \WT
contributions to the \dz distribution are shown in Figure~\ref{dz}-(a).

\subsection{Charge Asymmetry}
The measured same-sign dilepton charge asymmetry is defined as
\begin{equation}
A_{\ell\ell}(\Delta z) = \frac{N^{++}(\Delta z) - N^{--}(\Delta z)}{N^{++}(\Delta z) + N^{--}(\Delta z)},
\end{equation}
where $N^{\pm\pm}(\Delta z)$ are the true dilepton yields
as a function of  $\Delta z$.

\begin{figure}[!thb]
\includegraphics[width=17cm,clip]{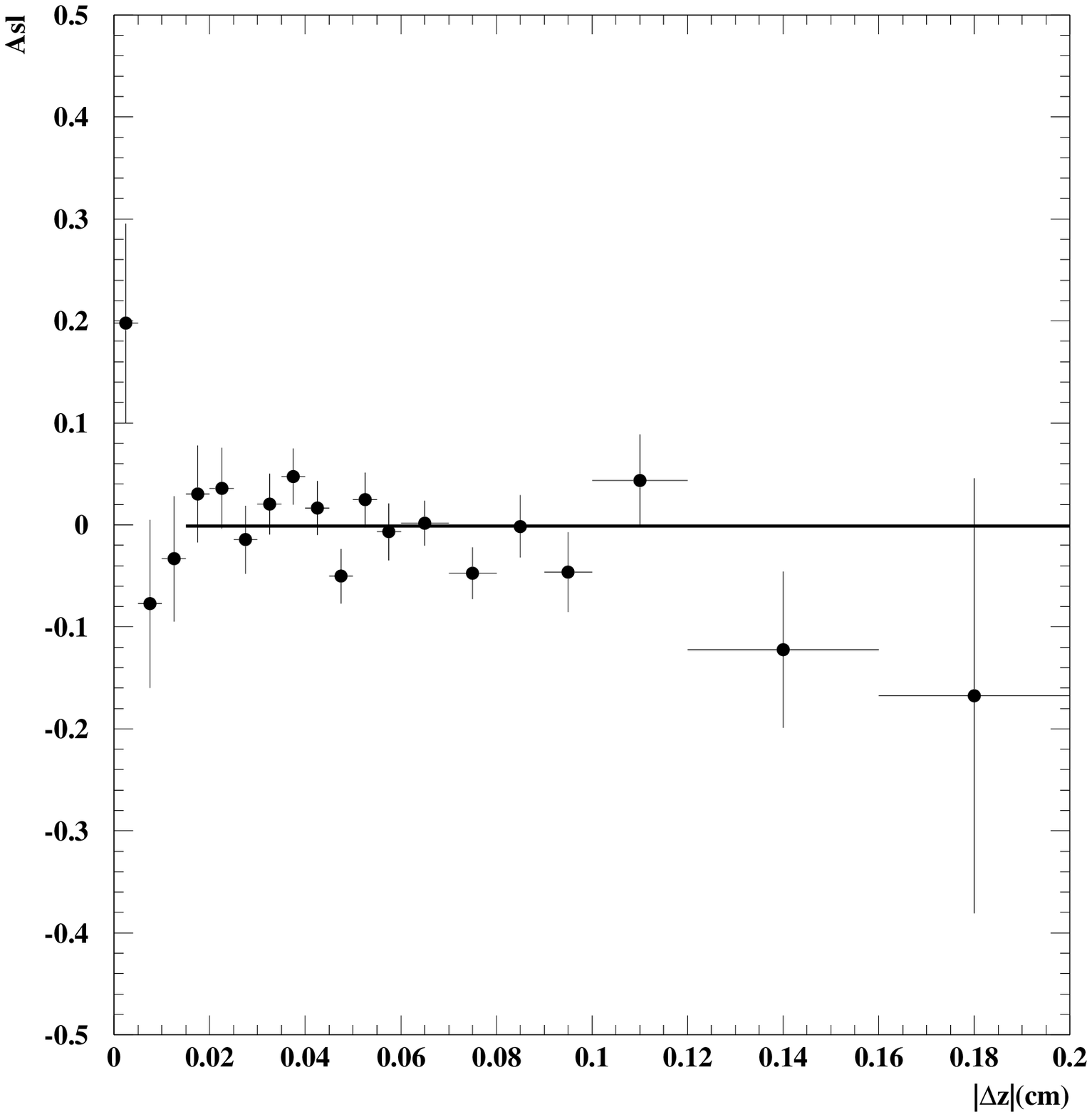}
\caption{$|\dz|$ distribution for $A_{\rm sl}$.}
\label{asl}
\end{figure}

Since $N^{\pm\pm}(\Delta z)$ are the sum of signal and background,
$N^{\pm\pm}(\Delta z) = N_s^{\pm\pm}(\Delta z) + N_b^{\pm\pm}(\Delta z)$,
the dilepton charge asymmetry  $A_{\rm sl}$ is related to $A_{\ell \ell}$
by 

\begin{equation}
A_{\ell\ell}(\Delta z) 
    = \frac{N_s^{++}(\dz) - N_s^{--}(\dz)}{N_s(\dz)} 
      \frac{N_s (\dz)}{( N_s (\dz)+ N_b (\dz))} 
    = A_{\rm sl}(\dz) d(\dz)
\label{fasl}
\end{equation}
where $N_s = N_s^{++} + N_s^{--}$ and $N_b = N_b^{++} + N_b^{--}$.
Here, $N_b^{++} = N_b^{--}$ is assumed.
A dilution factor, $d(\dz) = N_s (\dz)/ (N_s (\dz) + N_b(\dz))$, is
calculated using the signal and background yields, which are determined
in the fit given in Figure~\ref{dz}-(a). The result for $A_{\rm sl}(\dz)$,
determined from the measured $A_{\ell\ell}(\Delta z)$ and
$d(\Delta z)$, is shown in Figure~\ref{asl}.

The dilepton charge asymmetry $A_{\rm sl}$ is a time integrated
quantity and does not depend on \dz.
Fitting this distribution to a constant in the region of
$0.015~\cm<|\Delta z|<0.200~\cm$, yields
$A_{\rm sl} = (-1.1 \pm 7.9)\times 10^{-3}$, with a $\chi^2/n.d.f.$
of 36.20/36 (again note that only statistical errors
are included).
The optimum fitting range is determined using a MC study that maximizes
$N_{s}/\sqrt{N_{s} + N_{b}}$.

\subsection{Cross Checks}
As a consistency check, $A_{\rm sl}$ is obtained separately
for the $ee$, $\mu \mu$, and $e\mu$ data samples.
The results, $A_{\rm sl}(ee) = (-13.3 \pm 14.4)\times 10^{-3}$,
$A_{\rm sl}(\mu \mu) = (+16.5 \pm 17.3)\times 10^{-3}$,
and $A_{\rm sl}(e\mu) =(-3.6 \pm 11.3)\times 10^{-3}$,
are consistent with the primary result.
Here the errors are statistical only.  

The validity of the assumption,
to extract $A_{\rm sl}$ from $A_{\ell\ell}(\dz)$,
is confirmed by repeating the fit without it. This yields
$N_b^{++} = 20200 \pm 212$ and $N_b^{--} = 19766 \pm
210$ in the range $0.015~\cm < |\dz| < 0.200~\cm$, which is consistent
with the initial assumption.

\subsection{Systematic Errors}
Systematic errors in the determination of $A_{\rm sl}$ come from uncertainties
in:
i) the event selection criteria;
ii) the continuum subtraction;
iii) corrections for efficiencies of track finding and
lepton identification and for lepton misidentification;
iv) the dilution factor determination; and
v) the \dz fit for the determination of $A_{\rm sl}$. 

Uncertainties in the selection criteria are estimated by repeating the
analysis with different threshold values. For the track selection, the
$\theta_\text{lab}$ lower (higher) limit is varied from the nominal
$30^{\circ}$ ($135^{\circ}$) to $40^{\circ}$ ($125^{\circ}$)
in ten (eighteen) steps, the closest-approach criterion in
$r$-$\phi$ from its nominal value of $0.05~\cm$ to $0.02~\cm$
in fourteen steps, the
closest-approach criterion in $r$-$z$ from $2~\cm$ to $1~\cm$
in fourteen steps,
the $p^*$ lower (higher) limit from its nominal value of
$1.2~\gevc$ ($2.3~\gevc$) to $1.3~\gevc$ ($2.2~\gevc$) in fourteen steps,
the requirement on the number of SVD hits by $\pm 1$
from its nominal value of greater than or equal to two for $r$-$z$ and one in 
$r$-$\phi$.
For the event selection, the
$\cos\theta^*_{\ell\ell}$ lower (higher) criterion is varied from nominal
$-0.80$ ($+0.95$) to $-0.60$ ($+0.70$) in twelve (seventeen)
steps.  In addition, the mass windows that reject \jpsi and
$\gamma \to e^+ e^-$ are widened by 20\% and 50\%, respectively.  

To estimate the systematic error from the continuum subtraction, the
analysis is repeated, varying the off-resonance yield by $\pm 1 \sigma$.

For systematic uncertainties arising from the corrections to lepton
candidates, we consider separately the effects due to the uncertainties in
each bin and possible correlated sources of error.  The fake rates from
pions, kaons and protons and the lepton identification efficiencies are
assumed to be independent.  The systematic uncertainties from these track
corrections are estimated by varying the efficiencies and fake rates by
$\pm 1 \sigma$ bin by bin.  The systematic effects associated with the
relative hadron multiplicities are estimated by varying the values by
$\pm 1 \sigma$ for all bins (both electron and muon) simultaneously.  The
uncertainties for each hadron species are estimated separately, assuming
that the multiplicities of each are uncorrelated.

The contribution from the detector \dz response function is estimated
by changing the response function shape according to the statistics of 
each $\Delta z$ bin of $\jpsi \to \ell^+\ell^-$ sample.
The contribution from the $69~\micron$ smearing is estimated by
repeating the analysis with $43~\micron$ and $96~\micron$ smearing, 
corresponding to the uncertainty in the difference of the resolution
between data and MC.
The contributions from uncertainties in \dm and \taub are
also estimated by varying the nominal values by $\pm 1\sigma$.
The fitting range of the dilution factor determination is varied
from its nominal range of $0.000~\cm < |\dz| < 0.200~\cm$
to $0.025~\cm < |\dz| < 0.200~\cm$.
For the fitting range for the determination of the final $A_{\rm sl}$, the
lower limit is varied from $0.000~\cm$ to $0.045~\cm$ in
nine steps.
The results of the systematic error determination for $A_{\rm sl}$ are
summarized in Table~\ref{sys}.

\begin{table}[htb]
  \begin{center}
    \caption{\label{sys}
      Source of systematic errors for the measurement of $A_{\rm sl}$.}
    \begin{tabular}{lll}
      \hline\hline
      Category   & Source   &  $\Delta A_{\rm sl}$ ($\times 10^{-3}$)\\
      \hline
      Event selection & Track selection & $\pm 2.61$\\
      & $\cos\theta^*_{\ell\ell}$ cut  & $\pm 0.63$\\
      & Lepton pair veto  & $\pm 2.33$\\
      \hline
      Continuum subtraction & & $\pm 4.88$\\
      \hline
      Track corrections & Track finding efficiency & $\pm 1.56$\\
      & Electron identification efficiency & $\pm 0.56$\\
      & Muon identification efficiency & $\pm 1.98$\\
      & Fake electrons & $\pm 0.45$\\
      & Fake muons & $\pm 0.81$\\
      & Relative multiplicity &  $\pm 0.56$\\
      \hline
      \dz fit for dileptons & Detector response function & $\pm 0.07$  \\
      & \dm & $\pm 0.08$ \\
      &\taub   & $\pm 0.07$  \\
      & $69~\micron$ smearing of background \dz & $\pm 0.13$ \\
      & Statistics of signal MC & $\pm 0.01$ \\
      & Statistics of background MC & $\pm 0.19$ \\
      & Fitting range  &  $\pm 0.04$\\
      &  Assuming $N_b^{++} = N_b^{--}$ & $\pm 1.59$ \\
      \hline
      \dz fit for $A_{\rm sl}$ & Fitting range & $\pm 1.30$ \\
      \hline
      Total  &   &  $\pm 6.97$ \\
      \hline
      \hline
    \end{tabular}
  \end{center}
\end{table}

\section{Conclusion}
The charge asymmetry for same-sign dilepton events from \ups decays has
been measured. The result is related to the $CP$ violation parameter in
\Bz-\BzB mixing, 
\begin{equation}
A_{\rm sl}=(-1.1 \pm 7.9(\text{stat}) \pm 7.0(\text{sys}))\times 10^{-3},
\end{equation}
or equivalently 
\begin{equation}
|q/p| =1.0005 \pm 0.0040(\text{stat}) \pm 0.0035(\text{sys}).
\end{equation}
The measured $A_{\rm sl}$ is
consistent with zero, or equivalently, $|q/p|$ is consistent with
unity. This implies that $CP$ violation in \Bz-\BzB mixing is below
the $O(10^{-2})$ level. The $CP$ violation parameter $\epsilon_\B$
can be calculated as 
\begin{equation}
\frac{\mathrm{Re}(\epsilon_B)}{1+|\epsilon_B|^2}
= (-0.3 \pm 2.0(\text{stat}) \pm1.7(\text{sys}))\times 10^{-3},
\end{equation}
using  the (exact) formula
\begin{equation}
\frac{\mathrm{Re}(\epsilon_B)}{1+|\epsilon_B|^2}
=0.5 \frac{1-\sqrt{(1-A_{\rm sl})/(1+A_{\rm sl})}}
          {1+\sqrt{(1-A_{\rm sl})/(1+A_{\rm sl})}}.
\end{equation}
These results are
consistent with previous measurements~\cite{epsilon} and
provide significantly more restrictive bounds.
Previous measurements of
$\mathrm{Re}(\epsilon_B)/(1+|\epsilon_B|^2)$
are listed in Table~\ref{comp} together with this measurement.

\begin{table}[htb]
  \begin{center}
    \caption{\label{comp}
      Comparison of the ${\mathrm{Re}(\epsilon_B)}/{(1+|\epsilon_B|^2)}$ measurements.}
    \begin{tabular}{lc}
      \hline\hline
      Experiment   &  ${\mathrm{Re}(\epsilon_B)}/{(1+|\epsilon_B|^2)}$\\
      \hline
      ALEPH & $-0.003 \pm 0.007$ \\
      CLEO  & $0.0035 \pm 0.0103 \pm 0.0015$ \\
      BABAR & ($1.2 \pm 2.9 \pm 3.6$)$\times 10^{-3}$ \\
      This experiment & ($-0.3 \pm 2.0 \pm 1.7$)$\times 10^{-3}$ \\
      \hline
      \hline
    \end{tabular}
  \end{center}
\end{table}

\section*{Acknowledgments}
We thank the KEKB group for the excellent operation of the
accelerator, the KEK cryogenics group for the efficient
operation of the solenoid, and the KEK computer group and
the National Institute of Informatics for valuable computing
and Super-SINET network support. We acknowledge support from
the Ministry of Education, Culture, Sports, Science, and
Technology of Japan and the Japan Society for the Promotion
of Science; the Australian Research Council and the
Australian Department of Education, Science and Training;
the National Science Foundation of China under contract
No.~10175071; the Department of Science and Technology of
India; the BK21 program of the Ministry of Education of
Korea and the CHEP SRC program of the Korea Science and
Engineering Foundation; the Polish State Committee for
Scientific Research under contract No.~2P03B 01324; the
Ministry of Science and Technology of the Russian
Federation; the Ministry of Higher Education, Science and Technology of the Republic of Slovenia;  the Swiss National Science Foundation; the National Science Council and
the Ministry of Education of Taiwan; and the U.S.\
Department of Energy.

\end{document}